\documentclass[aps,prb,twocolumn]{revtex4}

\usepackage{bm}
\usepackage{graphicx}

\begin{document}

\title{Cross-correlated shot noise in three-terminal superconducting hybrid nanostructures}
\author{Dmitry S. Golubev$^{1}$
and Andrei D. Zaikin$^{2,3}$
}
\affiliation{$^1$QTF Centre of Excellence, Department of Applied Physics, Aalto University, FI-00076 Aalto, Finland\\
$^2$Institute of Nanotechnology, Karlsruhe Institute of Technology (KIT), 76021 Karlsruhe, Germany\\
$^3$I.E.Tamm Department of Theoretical Physics, P.N.Lebedev Physical Institute, 119991 Moscow, Russia }

\begin{abstract}
We work out a unified theory describing both non-local electron transport and cross-correlated shot noise
in a three-terminal normal-superconducting-normal (NSN) hybrid nanostructure.  We describe noise cross correlations 
both for subgap and overgap bias voltages and for arbitrary distribution of channel transmissions in NS contacts.
We specifically address a physically important situation of diffusive 
contacts and demonstrate a non-trivial behavior of non-local shot noise  exhibiting both positive and negative cross correlations depending on the bias voltages. 
For this case, we derive a relatively simple analytical expression for cross-correlated noise power
which contains only experimentally accessible parameters.
\end{abstract}

\maketitle

\section{Introduction}

\begin{figure}[!ht]
\includegraphics[width=0.8\columnwidth]{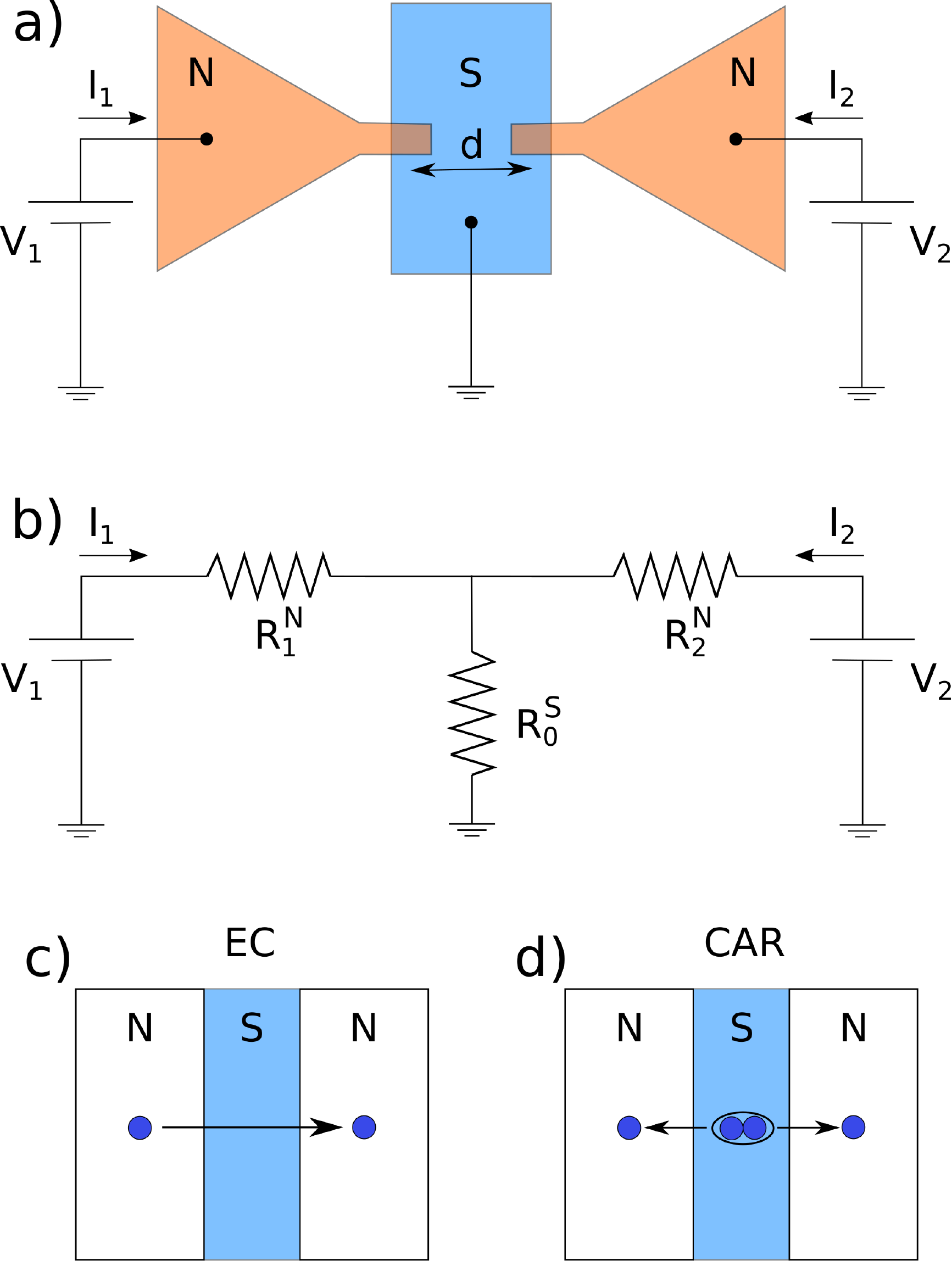}
\caption{(a) Schematics of an NSN structure under consideration.
The contacts between the normal leads and the superconductor (located at a distance $d$ from each other) may be described by an arbitrary distribution of channel transmissions or, else, may have a shape of short diffusive wires.
(b) Equivalent circuit of the same system in the normal state. Here $R_1^{N}$ and $R_2^N$ are the junction resistances and 
$R_0^S$ is the normal state resistance of the superconducting lead.
(c) Schematics of elastic cotunneling (EC) process in which an electron is transferred from one normal terminal 
to another one through an effective barrier formed by a superconductor.  
(d) Schematics of crossed Andreev reflection (CAR) process corresponding to splitting of a Cooper pair into two entangled electrons located in two normal terminals.}
\label{NISIN}
\end{figure}

It is well known that a normal metal attached to a superconductor also acquires superconducting properties. At low enough
temperatures proximity induced superconducting correlations may spread at long distances inside the normal metal leading to a wealth of interesting phenomena \cite{Bel}. Furthermore, electrons in two different normal metals may become coherent provided these metals are connected through a superconducting island with effective thickness $d$ shorter than the superconducting coherence length $\xi$. This effect has to do with the phenomenon of the so-called crossed Andreev reflection (CAR), in which a Cooper pair may split into two electrons going in two different normal leads \cite{DF}, see Fig. \ref{NISIN}d. 
This Cooper pair splitting process may be used to generate pairs of entangled electrons in different metallic electrodes \cite{Lesovik,Buttiker,Brange}, i.e. to experimentally realize a quantum phenomenon that could be of crucial importance for developing quantum communication technologies.

Crossed Andreev reflection is a quantum coherent process, which strongly affects electron transport in three-terminal normal metal - superconductor - normal metal  (NSN) 
hybrid structures at sufficiently low temperatures. This issue triggered a substantial theoretical \cite{FFH,BG,KZ1,KZ2,Belzig,GZ07,LY,GKZ} (see also further references therein) and experimental \cite{Beck1,Teun,Venkat1,Hof,Saclay,Basel3,Basel,Beck2,Beck3} interest over past years and is presently quite well understood.

Consider, e.g., an NSN structure depicted in Fig. \ref{NISIN}a. Applying bias voltages $V_1$ ad $V_2$ to two normal metallic electrodes and measuring electric currents $I_1$ and $I_2$ (depending on both voltages $V_1$ ad $V_2$)
it becomes possible to identify the contribution of CAR to non-local electron transport in such a structure. 
In fact, CAR is not the only process which contributes to the non-local transport in this case.
It competes with the so-called elastic cotunneling (EC), which does not produce entangled electrons. In the course of the latter process an electron is being transferred 
from one normal metal to another overcoming the effective barrier created by the energy gap inside the superconductor (see Fig. \ref{NISIN}c). In the zero temperature limit EC and CAR contributions to the low bias non-local conductance $\partial I_1/\partial V_2$ cancel each other in NSN structures with low transparency contacts\cite{FFH}.

One possible way to discriminate between CAR and EC processes 
is to investigate fluctuations of the currents $I_1$ and $I_2$. It is well known that in normal (i.e. non-superconducting) multiterminal structures cross correlations of current noise in different terminals are always negative due to the Pauli exclusion principle for electrons \cite{BB}.  In the presence of superconductivity
such cross correlations may become positive due to CAR. Hence, by measuring cross-correlated current noise in 
a system like the one depicted in Fig. \ref{NISIN}a it is possible to provide a clear experimental 
evidence for the presence of CAR in the system. 

A theoretical treatment of cross-correlated non-local current noise in NSN structures was pioneered in the works \cite{Belzig3,Hekking}
for the case of tunnel barriers at NS interfaces, and in the work \cite{Samuelsson} for a chaotic cavity coupled to normal and superconducting electrodes. 
This treatment indeed demonstrated that at certain voltage bias values CAR can dominate non-local shot noise giving rise to positive cross correlations. Later on theoretical analysis of noise cross correlations was extended to the case of arbitrary barrier transmissions  \cite{Melin2,Belzig2,GZ,Melin,Melin3,Ostrove}. In particular, for fully open barriers and at low enough temperature positive cross correlations were predicted to occur at any non-zero voltage bias values \cite{GZ}. Positively cross-correlated non-local shot noise was also observed in several  experiments \cite{Ch,Das}.

In this work we extend the existing theory of non-local shot noise in NSN hybrids\cite{GZ}, developed for non-interacting electrons, in at least two important aspects.
Firstly, here we relax the assumption \cite{GZ} restricting the energy to subgap values and develop the analysis
of both non-local electron transport and non-local shot noise at any voltage bias values $V_{1,2}$ and temperature $T$
both below and above the superconducting gap $\Delta$. Secondly, we do not anymore assume (unlike it was done in \cite{GZ}) that transmission probabilities for all conducting channels in the junction are equal and allow for an arbitrary transmission distribution. 
Following the analysis \cite{GZ}, we perform the lowest order expansion in the small ratio between the normal state resistance of the superconducting 
lead $R_0^S$ and the interface resistances  $R_{1,2}^N$ (see Fig. \ref{NISIN}b), which allows us to perform disorder averaging in a superconducting terminal exactly.
In this way, we derive a general
analytical expression for the cross-correlated non-local noise in the two contacts (\ref{S12full}).
We specifically address an important case of diffusive contacts, where the expression for the noise (\ref{S12}) greatly simplifies  and contains only experimentally accessible parameters.

The structure of the paper is as follows. In Sec. II  we derive a general expression for the cumulant generating function in an NSN structure with arbitrary distribution of conducting channel transmissions. In Sec. III we briefly recollect the results for both
local electron transport and local shot noise in a single NS contact thereby preparing our subsequent consideration of non-local effects. Non-local transport and non-local shot noise are addressed in details in Sec. IV paying special attention to an important physical situation of diffusive NS junctions. A couple of general and rather lengthy results are relegated to Appendix.

\section{Cumulant generating function}

In what follows we will consider an NSN structure depicted in Fig. \ref{NISIN}a. 
Normal metallic leads are connected to a bulk superconductor by two junctions characterized by a set of transmission probabilities $\tau_{1,n}$ and $\tau_{2,n}$, where
$n$ is the integer number enumerating all conducting channels.  The two junctions are located at a distance $d$ from each other which is assumed to be shorter that the superconducting coherence length $\xi$. 

Let $P_t(N_1,N_2)$ be the probability for $N_1$ and $N_2$ electrons to be transferred respectively through the junctions 1 and 2 during the observation time $t$. It is instructive to introduce the so-called cumulant generating function (CGF) ${\cal F}(\chi_1,\chi_2)$
by means of the relation
\begin{eqnarray}
e^{{\cal F}(\chi_1,\chi_2)} = \sum_{N_1,N_2} e^{-iN_1\chi_1-iN_2\chi_2}P_t(N_1,N_2).
\label{CGF1}
\end{eqnarray}   
The parameters $\chi_1$ and $\chi_2$ are denoted as counting fields.
The average currents through the junctions $I_r=\langle \hat I_r(t)\rangle$,
and the correlation functions of the currents, 
\begin{eqnarray}
S_{rr'} &=& \frac{1}{2}\int dt\big[ \big\langle  \hat I_r(t_0+t) \hat I_{r'}(t_0) +  \hat I_{r'}(t_0) \hat I_r(t_0+t) \big\rangle
\nonumber\\ &&
-\, 2\langle I_r(t_0)\big\rangle \langle I_{r'}(t_0)\big\rangle\big]
\end{eqnarray}
are expressed via the CGF as follows
\begin{eqnarray}
I_r &=& \lim_{t\to 0}\frac{ie}{t}\frac{\partial {\cal F}}{\partial\chi_r}\bigg|_{\chi_r=0},
\label{IrCGF}
\nonumber\\
S_{rr'} &=& -\lim_{t\to 0}\frac{e^2}{t}\frac{\partial^2 {\cal F}}{\partial\chi_r\partial\chi_{r'}}\bigg|_{\chi_r=0}.
\label{SCGF}
\end{eqnarray} 

In order to evaluate the CGF for the system depicted in Fig. 1 we will make use of the effective action approach \cite{GZ}.
The Hamiltonian of our system is expressed in the form
\begin{eqnarray}
H=H_1+H_2+H_S+H_{T,1}+H_{T,2},
\end{eqnarray}
where  $H_{1,2}$ are the Hamiltonians of the normal leads,
\begin{eqnarray}
H_r=\sum_{\alpha=\uparrow,\downarrow}\int d{\bm x}\
\hat \psi^\dagger_{r,\alpha}({\bm x})\left(-\frac{\nabla^2}{2m}-\mu-eV_r\right)\hat \psi_{r,\alpha}({\bm x}),
\end{eqnarray}
$\hat \psi^\dagger_{r,\alpha}({\bm x}),\hat \psi_{r,\alpha}({\bm x})$ 
are the creation and annihilation operators for an electron with a spin projection $\alpha$ at a point ${\bm x}$,
$m$ is the electron mass, $\mu$ is the chemical potential, $V_r$ is the electric potential applied to the lead $r$,
\begin{eqnarray}
H_S &=& \int d{\bm x} \bigg[
\sum_{\alpha}\hat\psi^\dagger_{S,\alpha}({\bm x})\left(-\frac{\nabla^2}{2m}-\mu+U_{\rm dis}({\bm x})\right)\hat\psi_{S,\alpha}({\bm x})
\nonumber\\ &&
+\,\Delta \hat\psi^\dagger_{S,\uparrow}({\bm x})\hat\psi^\dagger_{S,\downarrow}({\bm x})
+ \Delta^* \hat\psi_{S,\downarrow}({\bm x})\hat\psi_{S,\uparrow}({\bm x})
\bigg]
\end{eqnarray}
is the Hamiltonian of a superconducting electrode with the order parameter $\Delta$ and disorder potential $U_{\rm dis}({\bm x})$, and the terms
\begin{eqnarray}
H_{T,r}&=&\int_{{\cal A}_r} d^2{\bm x}\sum_{\alpha=\uparrow,\downarrow}
\big[ t_r({\bm x})  \hat\psi^\dagger_{r,\alpha}({\bm x}) \hat\psi_{S,\alpha}({\bm x})
\nonumber\\ &&
+\, t_r^*({\bm x})  \hat\psi_{S,\alpha}^\dagger({\bm x})\hat\psi_{r,\alpha}({\bm x})  \big]
\label{HT}
\end{eqnarray}
describe electron transfer through the contacts between the superconductor and the normal leads.
In Eq. (\ref{HT}) the surface integrals run over the contact areas ${\cal A}_r$,  and $t_r({\bm x})$ are the coordinate dependent tunneling amplitudes. Note that here we do not consider the case of spin active interfaces \cite{KZ2}, hence the amplitudes $t_r({\bm x})$ do not depend on the spin projection. 

One can introduce the wave functions in the leads corresponding to incoming and outgoing scattering states 
in the $n-$th conducting channel of the $r-$th junction, $\psi_{r,n}({\bm x})$, and
expand the electronic operators as
\begin{eqnarray}
\hat\psi_{r,\alpha}({\bm x}) &=& \sum_n \psi_{r,n}({\bm x}) \hat a_{n,\alpha},
\nonumber\\
\hat\psi_{S,\alpha}({\bm x}) &=& \sum_n \psi_{S,n}({\bm x})\hat c_{n,\alpha}.
\label{psi_rn}
\end{eqnarray}
The Hamiltonians (\ref{HT}) then acquire the form
\begin{eqnarray}
H_{T,r} = \sum_{\alpha=\uparrow,\downarrow} \sum_{n} [ t_{r,n}  \hat a^\dagger_{n,\alpha}\hat c_{n,\alpha} + t_{r,n}^*  \hat c^\dagger_{n,\alpha}\hat a_{n,\alpha}  ],
\end{eqnarray}
where $t_{r,n}=\int_{{\cal A}_r} d^2{\bm x}\psi_{S,n}^*({\bm x})t_{r}({\bm x})\psi_{r,n}({\bm x})$ are the matrix elements of the tunneling amplitude. These matrix elements are related to the channel transmission probabilities $\tau_{r,n}$ by means of the standard relation \cite{Carlos}
\begin{equation}
\tau_{r,n} = {4\alpha_{r,n}}/{(1+\alpha_{r,n})^2},
\label{ttr}
\end{equation}
with $\alpha_{r,n}=\pi^2\nu_r\nu_S|t_{r,n}|^2$ and
$\nu_j$  being the density of states in the corresponding electrode (here $j=1,2,S$). 

The CGF (\ref{CGF1}) can formally be expressed as
\begin{eqnarray}
{\cal F}
=\ln\left[ {\rm tr}\left( e^{-i\chi_1 \hat N_1-i\chi_2 \hat N_2} e^{-i H t} e^{i\chi_1 \hat N_1+i\chi_2 \hat N_2} 
\hat\rho_0 e^{i H t}\right) \right],
\nonumber\\
\label{FCS1}
\end{eqnarray}
where $\hat N_r = \sum_{\alpha=\uparrow,\downarrow}\sum_n \hat a_{\alpha,n}^\dagger \hat a_{\alpha,n}$ are the electron number operators in the normal leads and $\hat\rho_0=\exp(-H/T)/{\rm tr}[\exp(-H/T)]$ is the equilibrium density matrix of the system. 
The above expression can identically be transformed to 
\begin{eqnarray}
{\cal F} = \ln\left[ {\rm tr}\left( e^{-i H(\chi_1,\chi_2) t} 
\hat\rho_0 e^{i H(-\chi_1,-\chi_2) t}\right) \right],
\label{FCS}
\end{eqnarray}
where 
\begin{eqnarray}
&& H(\chi_1,\chi_2)= e^{-i(\chi_1\hat N_1+\chi_2\hat N_2)/2} H e^{i(\chi_1\hat N_1+\chi_2\hat N_2)/2}
\nonumber\\ &&
=\, H_1+H_2+H_S+H_{T,1}(\chi_1)+H_{T,2}(\chi_2)
\end{eqnarray}
and
\begin{eqnarray}
&& H_{T,r}(\chi) = e^{-i\chi_r\hat N_r/2} H_{T,r} e^{i\chi_r\hat N_r/2}
\nonumber\\ &&
=\,\sum_{\alpha=\uparrow,\downarrow} \sum_{n} [t_{r,n} e^{i\chi_r/2} \hat c^\dagger_{n,\alpha}\hat a_{n,\alpha} 
+ t_{r,n}^* e^{-i\chi_r/2} \hat a^\dagger_{n,\alpha}\hat c_{n,\alpha} ].
\nonumber
\end{eqnarray}

The CGF (\ref{FCS}) can be evaluated in a straightforward manner with the aid of the path integral technique \cite{GZ} which yields
\begin{eqnarray}
{\cal F}(\chi_1,\chi_2) = {\rm tr}\,[\ln\hat {\cal G}^{-1}(\chi_1,\chi_2) ],
\label{FCS2}
\end{eqnarray}
where $\hat {\cal G}^{-1}$ is the Keldysh Green function of our system
\begin{eqnarray}
{\cal G}^{-1}(\chi_1,\chi_2)=\left(
\begin{array}{ccc}
\check G_1^{-1} & \check t_1(\chi_1) & 0 \\
\check t_1^\dagger(\chi_1) & \check G_S^{-1} & \check t_2(\chi_2) \\
0 & \check t_2^\dagger(\chi_2) & \check G_2^{-1}
\end{array}\right),
\end{eqnarray}
the $4\times 4$ matrices $\check G_{j}^{-1}$ represent the inverse Keldysh Green functions 
of isolated normal and superconducting leads and $\check t_r$ is the diagonal $4\times 4$ matrix
in the Nambu - Keldysh space describing tunneling between the leads,
\begin{eqnarray}
\check t_r(\chi_r) = \left(
\begin{array}{cccc}
-t_r e^{-i\frac{\chi_r}{2}} & 0 & 0 & 0 \\
0 & t_r e^{i\frac{\chi_r}{2}} & 0 & 0 \\
0 & 0 & t_r e^{i\frac{\chi_r}{2}} & 0 \\
0 & 0 & 0 & -t_r e^{-i\frac{\chi_r}{2}}
\end{array}
\right).
\label{t}
\end{eqnarray}
The CGF (\ref{FCS2}) can be cast to the form
\begin{eqnarray}
{\cal F} = {\rm tr}\ln\big[ \check I - \check t_1^\dagger \check G_1\check t_1\check G_S
- \check t_2^\dagger \check G_2\check t_2 \check G_S \big].
\label{S}
\end{eqnarray}
where $\check I$ is the unity operator.

The Fourier transformed Green function of a superconducting island, $\check G_S(E)=\int dt\, e^{iE(t-t')}\check G_S(t-t',{\bm x},{\bm x}')$, reads
\begin{eqnarray}
\check G_S(E) &=& \frac{\hat G_{R}(E) +\hat G_{A}(E)}{2}\otimes\hat\sigma_z 
\nonumber\\ &&
+\, \frac{\hat G_{R}(E) - \hat G_{A}(E)}{2}\otimes\hat Q_S(E)\hat\sigma_z.
\label{GS}
\end{eqnarray}
Here
\begin{eqnarray}
&& \hat G_{R,A}(E)=
\left(\begin{array}{cc} G_{R,A}(E,{\bm x},{\bm x}') & F^+_{R,A}(E,{\bm x},{\bm x}') \\
F_{R,A}(E,{\bm x},{\bm x}')  & G^+_{R,A}(E,{\bm x},{\bm x}') \end{array}\right)
\nonumber\\ &&
=\,\sum_n\frac{\varphi_{n}({\bm x})\varphi_{n}^*({\bm x}')}{(E\pm i0)^2-\xi_{n}^2-|\Delta|^2}
\left(\begin{array}{cc} E+\xi_{n} & \Delta^* \\ \Delta & E-\xi_{n} \end{array}\right),
\nonumber\\
\label{GRA}
\end{eqnarray} 
are retarded and advanced Green functions and the matrix
\begin{eqnarray}
\hat Q_S(E)=\left(\begin{array}{cc} 1-2n_S(E) & 2n_S(E) \\ 2-2n_S(E) & 2n_S(E)-1 \end{array}\right)
\label{QS}
\end{eqnarray}
depends on the quasiparticle distribution function  $n_S(E)$ in a superconductor and has the property $\hat Q^2(E)=1$.
The wave functions $\varphi_n({\bm x})$, appearing in Eq. (\ref{GRA}) are the eigenfunctions
of a single electron Hamiltonian of the superconducting lead with eigenenergies $\xi_n$, i.e. they are the solutions
of the Schr\"odinger equation
\begin{eqnarray}
\left(-\frac{\nabla^2}{2m}-\mu+U_{\rm dis}({\bm x})\right)\varphi_{n}({\bm x}) = \xi_n \varphi_{n}({\bm x}).
\end{eqnarray}
Note that the wave functions $\varphi_{n}({\bm x})$ differ from the functions $\psi_{r,n}({\bm x})$ introduced earlier
in Eqs. (\ref{psi_rn}).
The expressions for the Green functions in the normal leads are recovered from Eqs. (\ref{GS})-(\ref{QS}) by replacing
$S \to r=1,2$ and setting $\Delta=0$.

Following the analysis \cite{GZ} let us define the self-energies $\check\Sigma_r(\chi_r) = \check t^\dagger_r \check G_r\check t_r$ and derive their matrix elements in the basis of the  scattering states wave functions in the corresponding contact. We obtain
\begin{eqnarray}
&& \check\Sigma_r^{mn}(\chi_r,E)= \int_{{\cal A}_r} d^2{\bm x}d^2{\bm x}'\,\psi_{S,m}^*({\bm x}) 
\check t^\dagger_r(\chi_r,{\bm x})
\nonumber\\ && \times\,
\check G_r(E,{\bm x},{\bm x}')\check t_r(\chi_r,{\bm x}')\psi_{S,n}({\bm x}')
\nonumber\\ &&
=\,\frac{\alpha_{r,n}}{\pi i \nu_S}\delta_{mn} 
\left(\begin{array}{c} 
\hat\sigma_z e^{-i\frac{\hat\sigma_z\chi}{2}}\hat Q(E-eV_r)e^{i\frac{\hat\sigma_z\chi}{2}} \qquad 0 \\ 
0 \qquad \hat\sigma_z e^{i\frac{\hat\sigma_z\chi}{2}}\hat Q(E+eV_r)e^{-i\frac{\hat\sigma_z\chi}{2}} \end{array}\right),
\nonumber\\
\label{Sigma}
\end{eqnarray}
where the matrices $\hat Q_r(E)$ are defined in the same way as in Eq. (\ref{QS}), i.e.
\begin{eqnarray}
\hat Q_r(E)=\left(\begin{array}{cc} 1-2n_r(E) & 2n_r(E) \\ 2-2n_r(E) & 2n_r(E)-1 \end{array}\right),
\end{eqnarray}
and $n_r(E)$ are the distribution functions of electrons in the normal leads.
Note that by performing a proper rotation in the basis of the scattering wave functions in the superconductor one can always diagonalize the self-energies $\check\Sigma_r^{mn} \propto \delta_{mn}$. Hence, the CGF (\ref{S}) can be expressed in the form
\begin{eqnarray}
{\cal F}(\chi_1,\chi_2) 
={\rm tr}\,\ln\big[ \check I -\check\Sigma_1(\chi_1) \check G_S - \check\Sigma_2(\chi_2) \check G_S \big].
\label{CGF3}
\end{eqnarray}

Unfortunately, the CGF (\ref{CGF3}) cannot be evaluated exactly. In order to proceed and to account for the effects of CAR we carry out a perturbative expansion of the CGF (\ref{CGF3}) 
in powers of the "off-diagonal" component of the superconductor Green function  $\check G_S({\bm x},{\bm x}')$, in which
the points ${\bm x}$ and ${\bm x}'$ belong to different junctions. This expansion is justified provided
the normal state resistance of the superconducting lead $R_0^S$ remains small as compared to the 
contact resistances $R_{1}^N,R_2^N$, and it is essentially equivalent to linearizing the Usadel equation.
The latter simplification is routinely performed \cite{Volkov1} in order to fully analytically describe various non-trivial non-equilibrium effects in superconducting hybrid structures, such as, e.g., the sign inversion of the Josephson critical current  in SNS-like junctions \cite{Morpurgo,Volkov2}.
To this end, we define the operator $\check A=\check\Sigma_1(\chi_1) \check G_S + \check\Sigma_2(\chi_2) \check G_S$
and formally rewrite the expression (\ref{CGF3}) in the form
\begin{eqnarray}
{\cal F}(\chi_1,\chi_2)
={\rm tr}\,\ln\left[\left( \begin{array}{cc} \check I_{11} -\check A_{11} & -\check A_{12} \\ -\check A_{21} & \check I_{22} - \check A_{22} \end{array}\right)\right],
\label{CGF}
\end{eqnarray}
where the subscripts indicate the contact at which the coordinates ${\bm x}$ 
(first index) or ${\bm x}'$ (second index) are located.
Expanding in the small "off-diagonal" components $\check A_{12}, \check A_{21}$ to the lowest non-vanishing order, we arrive at the result
\begin{eqnarray}
{\cal F}(\chi_1,\chi_2) = {\cal F}_1(\chi_1) + {\cal F}_2(\chi_2) + {\cal F}_{12}(\chi_1,\chi_2),
\end{eqnarray}
where 
\begin{eqnarray}
{\cal F}_r(\chi_r) = {\rm tr}\,\ln\big[ \check I -\check\Sigma_r(\chi_r) \check G_{S,rr} \big],\;\; r=1,2
\label{Fr}
\end{eqnarray}
are the local contributions and the term
\begin{eqnarray}
{\cal F}_{12}(\chi_1,\chi_2) &=& 
-t\int \frac{dE}{2\pi}{\rm tr}\left[ (\check I_{11} -\check\Sigma_1(\chi_1) \check G_{S,11})^{-1}\check A_{12}
\right.
\nonumber\\ && \times\,
\left.
(\check I_{22} -\check\Sigma_2(\chi_2) \check G_{S,22})^{-1}\check A_{21} \right]
\label{F12}
\end{eqnarray}
accounts for non-local effects. Note that in Eq. (\ref{F12}) we replaced the double time integration by a single integral over energy which is appropriate in the long time limit.

The expressions (\ref{Fr}) and (\ref{F12}) contain the Green functions of the superconductor $G_{R,A}$,
which oscillate at the scale of the Fermi wavelength. One can simplify these expressions by averaging over disorder. 
Such averaging can be handled with the aid of the following relations \cite{Brouwer}:
\begin{eqnarray}
&& \sum_n \langle\varphi_n({\bm x})\varphi_n^*({\bm x}')\rangle\delta(\xi-\xi_n) = \nu_S w(|{\bm x}-{\bm x}'|),
\label{av1}
\\
&&\sum_{mn}\langle \varphi_n({\bm x})\varphi_n^*({\bm x}')\varphi_m({\bm x}')\varphi_m^*({\bm x})\rangle\delta(\xi-\xi_n)\delta(\xi'-\xi_m) 
\nonumber\\ &&
=\,\nu_S^2 w^2(|{\bm x}-{\bm x}'|) + \frac{\nu_S}{\pi}\,{\rm Re}\, D(\xi-\xi',{\bm x},{\bm x}')
\nonumber\\ &&
+\, \frac{\nu_S}{\pi} w^2(|{\bm x}-{\bm x}'|) \,{\rm Re}\,C\left( \xi-\xi',\frac{{\bm x}+{\bm x}'}{2},\frac{{\bm x}+{\bm x}'}{2} \right).
\label{av}
\end{eqnarray}
Here $w(r)=e^{-r/2l_e}\sin(k_Fr)/k_Fr$, $l_e$ is the mean free path of electrons, and $D(\omega,{\bm x},{\bm x}')$, $C(\omega,{\bm x},{\bm x}')$ 
are, respectively, the diffuson and the Cooperon. 

In what follows we will assume that the distance between the two junctions
is shorter than the effective dephasing length for electrons, in which case the diffuson and the Cooperon coincide with each other, $D(\omega,{\bm x},{\bm x}')=C(\omega,{\bm x},{\bm x}')$, being determined by the fundamental solution of the diffusion equation
\begin{eqnarray}
(-i\omega - D_S\nabla_{{\bm x}}^2 ) D(\omega,{\bm x},{\bm x}')=\delta({\bm x}-{\bm x}'),
\end{eqnarray} 
where $D_S=v_Fl_e/3$ is the diffusion constant in the superconductor. 

Let us for simplicity ignore the influence of the proximity effect on local transport properties of the contacts
and replace the Green functions $\check G_{S,11}$ and $\check G_{S,22}$ appearing
in Eqs. (\ref{Fr}), (\ref{F12}) by their disorder averaged values $\langle\check G_{S,11}\rangle$ and $\langle\check G_{S,22}\rangle$. Averaging of pairwise products of the Green function components $\check G_{S,12}$ and $\check G_{S,21}$ (contained in the non-local terms $\check A_{12}$ and $\check A_{21}$ in Eq. (\ref{F12})) is carried out with the aid
of Eq. (\ref{av}). Further simplifications occur if we recall that the distance between the contacts $d$ remains shorter than the superconducting coherence length $\xi=\sqrt{D_S/2\Delta}$. In this case one can set $\omega=0$ in the argument of the diffuson, i.e. we replace $D(\omega,{\bm x},{\bm x}')\to D(0,{\bm x},{\bm x}')$. Finally, we also assume complete randomization
of the electron trajectories connecting the two contacts inside a disordered superconductor, meaning that an electron
leaving the junction 1 via the conduction channel $n$ has the same probability to arrive at the contact 2 
in any of its conduction channels.  In this way we bring the CGF (\ref{F12}) to the form 
\begin{eqnarray}
{\cal F}_{12} &=& t\frac{2e^2 R_0^S}{\pi}\sum_{n,m}\int dE \,{\rm tr}\left\{
\left(\check I_{11} - \check\Sigma_1(\chi_1)\langle\check G_{S,11}\rangle\right)_n^{-1} 
\right.
\nonumber\\ && \times\,
\left.
\alpha_{1,n}\check Q_1\left(\check I_{22} - \check\Sigma_2(\chi_2)\langle\check G_{S,22}\rangle\right)_m^{-1} \alpha_{2,m}\check Q_2
\right.
\nonumber\\ &&
\left.
-\, \left[1-\left(\check I_{11} - \check\Sigma_1(\chi_1)\langle\check G_{S,11}\rangle\right)_n^{-1}\right]
\right.
\nonumber\\ && \times\,
\left.
\left[1-\left(\check I_{22} - \check\Sigma_2(\chi_2)\langle\check G_{S,22}\rangle\right)_m^{-1}\right]
\right\},
\label{action}
\end{eqnarray}
where the sum runs over all conducting channels
of the contact 1 (index $n$) and of the contact 2 (index $m$),
\begin{eqnarray}
\check Q_r(E) = \left(\begin{array}{cc} \hat Q_r(E-eV_r) & 0 \\ 0 & -\hat Q_r(E+eV_r) \end{array}\right),
\end{eqnarray}
and $R_0^S$ is the characteristic resistance which sets the scale for non-local effects in our system. It is defined as
\begin{eqnarray}
R_0^S=\frac{1}{2e^2\nu_S {\cal A}_1{\cal A}_2}\int_{{\cal A}_1} d^2{\bm x}_1 \int_{{\cal A}_2} d^2{\bm x}_2\, D(0,{\bm x}_1,{\bm x}_2)
\end{eqnarray}
being approximately equal to the total resistance of the superconducting electrode measured in the normal state 
between the ground and the region to which the normal leads are attached, see Fig. \ref{NISIN}b. 

Equation (\ref{action}) for the non-local part of CGF represents the main result of this section which will be directly employed in our subsequent analysis.

\section{Local transport and noise in a single NS junction}

Before turning to non-local effects let us briefly recollect the well known results for both electron transport and noise
in single NS junctions  paying special attention to the case of a diffusive interface between the two metals.
Following a seminal work by Blonder, Tinkham and Klapwijk \cite{BTK} we define the
probabilities for scattering processes in the junction for every conducting channel. 
Specifically, these are the probabilities of Andreev reflection, $A_{r,n}$, of normal reflection, $B_{r,n}$, of 
normal transmission, $C_{r,n}$, and of the transmission with the conversion of an electron into a hole, $D_{r,n}$,
in the junction $r$.
At subgap energies $|E|<\Delta$ we have \cite{BTK}
\begin{eqnarray}
A_{r,n}(E) = \frac{4\alpha_{r,n}^2 \Delta^2}{(1+\alpha_{r,n}^2)^2\Delta^2 - (1-\alpha_{r,n}^2)^2 E^2},
\end{eqnarray}
$B_{r,n}=1-A_{r,n}$, $C_{r,n}=D_{r,n}=0$; while at $|E|>\Delta$ one gets\cite{BTK}
\begin{eqnarray}
A_{r,n}(E) &=& \frac{4\alpha_{r,n}^2 \left(N_S^2(E)-1\right)}{\left(1+\alpha_{r,n}^2 + 2\alpha_{r,n}N_S(E)\right)^2},
\nonumber\\
B_{r,n}(E) &=& \frac{(1-\alpha_{r,n}^2)^2}{\left(1+\alpha_{r,n}^2 + 2\alpha_{r,n}N_S(E)\right)^2},
\nonumber\\
C_{r,n}(E) &=& \frac{2\alpha_{r,n}(1+\alpha_{r,n})^2 \left(N_S(E)+1\right)}{\left(1+\alpha_{r,n}^2 + 2\alpha_{r,n}N_S(E)\right)^2},
\nonumber\\
D_{r,n}(E) &=& \frac{2\alpha_{r,n}(1-\alpha_{r,n})^2 \left(N_S(E)-1\right)}{\left(1+\alpha_{r,n}^2 + 2\alpha_{r,n}N_S(E)\right)^2},
\end{eqnarray}
where $N_S(E)=\theta(|E|-\Delta)|E|/\sqrt{E^2-\Delta^2}$ is the density of states in the superconductor,
and $\theta(x)$ is the Heaviside step function.

The CGF of a single contact (\ref{Fr}) can be evaluated exactly,  it is  presented in Appendix, see the Eq. (\ref{Floc}).
This result allows one to immediately reconstruct the well-known expression for the (local) current (\ref{IrCGF}) in the $r$-th junction \cite{BTK}
\begin{eqnarray}
I_r = \frac{e}{2\pi}\sum_n \int dE \, g(E,\alpha_{r,n}) (n_r^- - n_r^+).
\label{Ir0}
\end{eqnarray}
Here $n_j^-$ and $n_j^+$ are the distribution functions for respectively electrons and holes  in the normal leads,
\begin{eqnarray}
n_r^\pm = \frac{1}{1+e^{(E\pm eV_r)/T_r}},
\end{eqnarray}
and
\begin{eqnarray}
g(E,\alpha_{r,n}) = 2A_{r,n}(E) + C_{r,n}(E) + D_{r,n}(E)
\label{gE}
\end{eqnarray}
is the dimensionless spectral conductance in the $n-$th channel.

The expression for local current noise in the $r$-th junction, given by the derivative (\ref{SCGF}), 
\begin{eqnarray}
S_{rr} = \frac{1}{2}\int dt \left[\left\langle \hat I_r(t)\hat I_r(0) + \hat I_r(0)\hat I_r(t) \right\rangle - 2\left\langle\hat I_r(0)\right\rangle\right],  
\end{eqnarray}
is recovered analogously. We get \cite{MK,AD}
\begin{eqnarray}
&& S_{rr} = \frac{e^2}{2\pi}\sum_n\int dE 
\bigg[ 4\,\theta(\Delta-|E|)\,A_{r,n}(1-A_{r,n}) 
\nonumber\\ &&
\times\, w(n_r^-,n_r^+) + \left(C_{r,n}+D_{r,n}-(C_{r,n}-D_{r,n})^2\right)
\nonumber\\ && \times\,
(w(n_r^-,n_S)+w(n_r^+,n_S))
\nonumber\\ &&
+\, (2A_{r,n}+C_{r,n}+D_{r,n})^2\frac{w(n_r^-,n_r^-)+w(n_r^+,n_r^+)}{2} 
\nonumber\\ &&
+\,(C_{r,n}-D_{r,n})^2 w(n_S,n_S)\bigg].
\end{eqnarray}
Here we introduced the following combination of the distribution functions 
\begin{eqnarray}
w(n_r,n_{r'}) = n_r(1-n_{r'})+(1-n_{r})n_{r'}.
\end{eqnarray} 

Let us specify the above results in the important case of diffusive contacts.
Provided the contact has the form of a short diffusive wire with the Thouless energy exceeding
the superconducting gap $\Delta$,  the transmission probability distribution is determined by the Dorokhov's formula \cite{Dorokhov}
\begin{eqnarray}
P_r(\tau_r) = \frac{\pi}{2e^2 R_r^N} \frac{1}{\tau_r\sqrt{1-\tau_r}},\;\; r=1,2.
\label{Dorokhov}
\end{eqnarray}
Here $R_{1,2}^N$ are the resistances of diffusive contacts in the normal state. Introducing the dimensionless parameters $\alpha_r$ in a way (\ref{ttr}),  
which translates the distribution (\ref{Dorokhov}) to the form 
\begin{eqnarray}
P_r(\alpha_r) = \frac{\pi}{2e^2 R_r^N} \frac{1}{\alpha_r},
\end{eqnarray}
and replacing the sum over conducting channels in Eq. (\ref{Ir0})  by an integral,  $\sum_n \to \int_0^1 d\alpha_r P_r(\alpha_r)$, 
we arrive at the expression for the current through a diffusive junction between N- and S-metals
\begin{eqnarray}
I_r = \frac{1}{2e} \int dE\, G_r(E) (n_r^- - n_r^+).
\label{Ir1}
\end{eqnarray}
Here $G_r(E) = (1/R_r^N) f_1(E/\Delta)$ is the spectral conductance of the $r$-th short diffusive wire,
and the dimensionless function $f_1(x)$, defined as $f_1(x)=\int_0^1 d\alpha\, g(x\Delta,\alpha)/2\alpha$ (here $g(E,\alpha)$ is the conductance  (\ref{gE})), reads
\begin{eqnarray}
f_1(x) = \frac{1}{2}\left(  \frac{\theta(1-|x|)}{|x|} + \theta(|x|-1)  |x|\right) \ln\left|\frac{|x|+1}{|x|-1}\right|.
\label{f1}
\end{eqnarray}
The I-V curve defined by Eqs. (\ref{Ir1}), (\ref{f1}) is illustrated in Fig. 2a. 

\begin{figure}[!ht]
\includegraphics[width=\columnwidth]{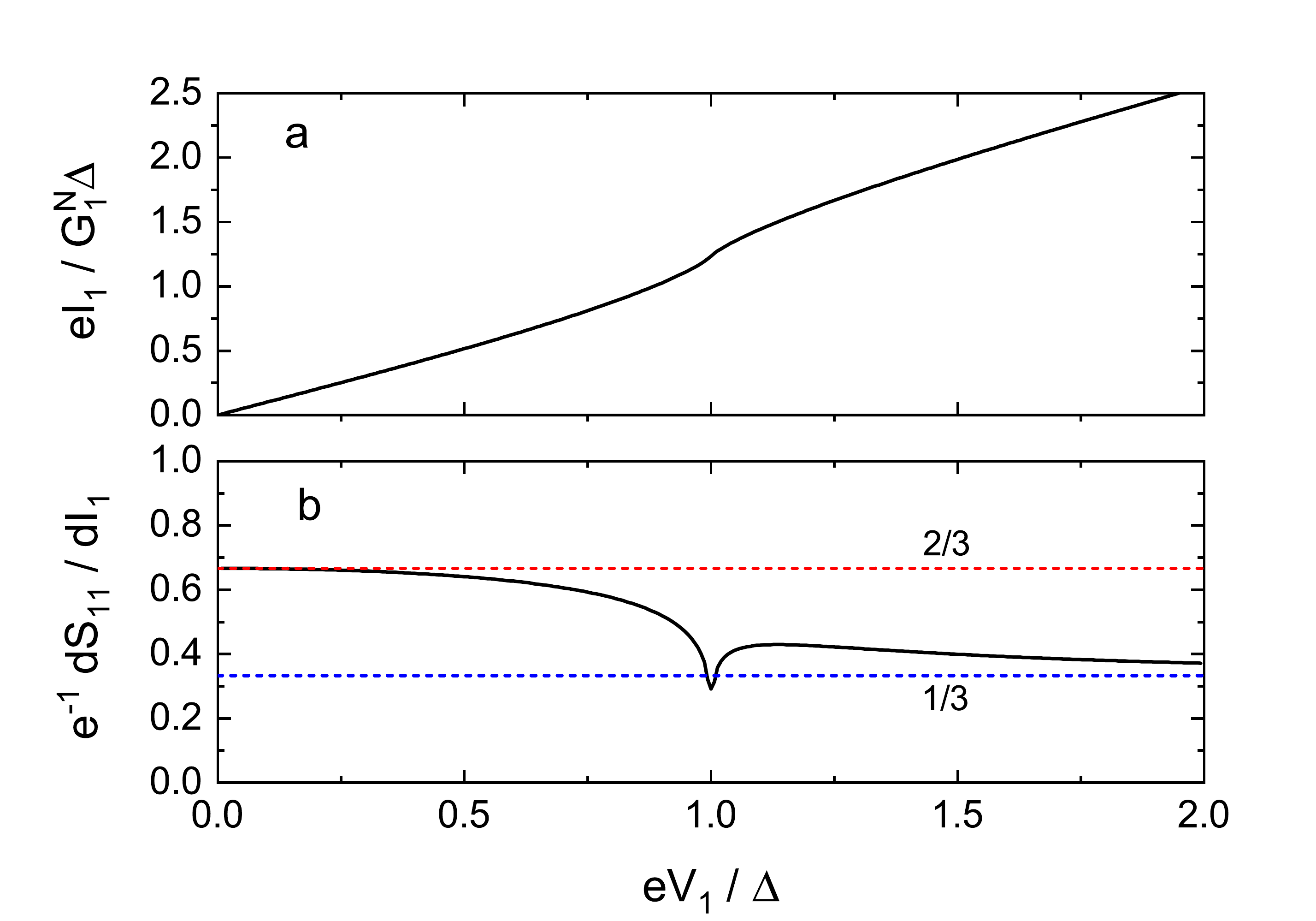}
\caption{The current $I_1$ (\ref{Ir1}) in a diffusive contact 1 (a) and the corresponding differential Fano factor 
$e^{-1}dS_{11}/dI_1$ (b) (determined from Eqs. (\ref{Ir1}), (\ref{Srr})) as functions of the voltage bias $V_1$ at $T=0$. Here we define $G_1^N \equiv 1/R_1^N$.}
\end{figure}

The local current noise power in $r$-th NS contact with a diffusive boundary between metals is constructed analogously. It reads
\begin{eqnarray}
&& S_{rr} = \frac{1}{R_r^N}\int_{|E|<\Delta} dE  \left[f_1\left(\frac{E}{\Delta}\right)-f_2\left(\frac{E}{\Delta}\right)\right]w(n_1^-,n_1^+) 
\nonumber\\ &&
+\, \frac{1}{2R_r^N} \int_{-\infty}^{+\infty} dE\, f_2\left(\frac{E}{\Delta}\right)\,\big[w(n_1^-,n_1^-)+w(n_1^+,n_1^+)\big] 
\nonumber\\ &&
+\,\frac{1}{2R_r^N}\int_{|E|>\Delta} dE
\bigg\{ \left[f_1\left(\frac{E}{\Delta}\right)-\left(2-\frac{\Delta^2}{E^2}\right)f_2\left(\frac{E}{\Delta}\right)\right]
\nonumber\\ && \times\,
\big[w(n_1^+,n_S)+w(n_1^-,n_S)\big]
\nonumber\\ &&
+\, 2f_2\left(\frac{E}{\Delta}\right) \left(1-\frac{\Delta^2}{E^2}\right)w(n_S,n_S) \bigg\},
\label{Srr}
\end{eqnarray}
where the dimensionless function $f_2(x)$, defined as $f_2(x)=\int_0^1 d\alpha\, g^2(x\Delta,\alpha)/4\alpha$, equals to
\begin{eqnarray}
f_2(x)&=&\frac{\theta(1-|x|)}{2x^2}\left(\frac{1+x^2}{2|x|}\ln\frac{1+|x|}{1-|x|}-1\right)
\nonumber\\ &&
+\, \theta(|x|-1)  \frac{x^2}{2}\left( |x|\ln\frac{|x|+1}{|x|-1} -2 \right).
\label{f2}
\end{eqnarray}
The differential Fano factor $e^{-1} dS_{11}/dI_1$ following from the above results is displayed in Fig. 2b.
At high voltage bias values it approaches the universal value $1/3$ expected for the normal metal, while at low bias $eV_1\ll \Delta$ the Fano factor becomes two times bigger due to the well-known charge doubling effect in the Andreev reflection regime \cite{Beenakker,Sanquer,Prober}.

At this stage we have completed our preparations and now can turn to a discussion of non-local effects.

\section{Non-local transport and noise in an NSN system}

To begin with, let us we evaluate the non-local correction to the current flowing through the contact $r$ due to the presence of another contat $r'$. We obtain
\begin{eqnarray}
I_r = \frac{1}{2}\int \left( \frac{e}{\pi}\sum_n g_{r,n}(E)\right)(n_r^- -n_r^+)
\nonumber\\ 
+\, \frac{1}{2e} \int dE\, G_{12}(E)(n_{r'}^- -n_{r'}^+),
\label{Ir2}
\end{eqnarray}
where the non-local spectral conductance $G_{12}(E)$ reads \cite{GKZ}
\begin{eqnarray}
G_{12}(E) = \frac{e^4R_0^S}{\pi^2}\sum_{n,m}\bigg[\theta(\Delta-|E|)\frac{\Delta^2-E^2}{\Delta^2}
\nonumber\\ 
+\,\theta(|E|-\Delta)\frac{E^2-\Delta^2}{E^2}\bigg]g_{r,n}(E)g_{r',m}(E).
\label{G1122}
\end{eqnarray}
Note that for simplicity in Eq. (\ref{Ir2}) we omitted disorder-induced corrections
to the local junction conductance  \cite{HN1,HN2,Tanaka} which are insignificant for our present discussion.

One can also work out a full analytical expression for the cross-correlated noise of the contacts $S_{12}$.
For the sake of completeness we present this rather lengthy expression in Appendix in Eq. (\ref{S12full}). 
In the important limit of low voltages and temperatures, $eV_{1,2},T\ll\Delta$ one can derive 
a simple analytical expression,
\begin{eqnarray}
&& S_{12} = G_{12}(0) \left[ -4(2-\beta_1-\beta_2)T - 4\beta_1 eV_1\coth\frac{eV_1}{T} 
\right.
\nonumber\\ &&
\left.
-\, 4\beta_2 eV_2\coth\frac{eV_2}{T}
+ \gamma_+ e(V_1+V_2) \coth\frac{e(V_1+V_2)}{2T} 
\right.
\nonumber\\ &&
\left.
-\, \gamma_- e(V_1-V_2) \coth\frac{e(V_1-V_2)}{2T} \right],
\label{S12subgap}
\end{eqnarray}
where
\begin{eqnarray}
\beta_r = \lim_{E\to 0}\frac{\sum_n A_{r,n}[1-A_{r,n}]}{\sum_n A_{r,n}}
\label{beta}
\end{eqnarray}
are the effective Fano factors of the junctions in the regime where Andreev reflection dominates the transport properties,
and the parameters $\gamma_\pm$ are defined as
\begin{eqnarray}
\gamma_\pm = \frac{\sum_{n,m} A_{1,n}A_{2,m}\left[ \frac{1-2A_{1,n}-2A_{2,m}+4A_{1,n}A_{2,m}}{\sqrt{A_{1,n}A_{2,m}}} \pm 1  \right] }
{\sum_{n,m} A_{1,n}A_{2,m}}.
\label{gamma}
\end{eqnarray}
Here the limit $E\to 0$ should be taken in the same way as in Eq. (\ref{beta}).

Eqs. (\ref{S12subgap})-(\ref{gamma}) constitute an important generalization of our previous result \cite{GZ}, where the assumption about equal transmissions of all conducting channels has been made. This assumption is lifted here, thus allowing
one to analyze the results for a variety of transmission distributions in the contacts.

\begin{figure}
\includegraphics[width=9cm]{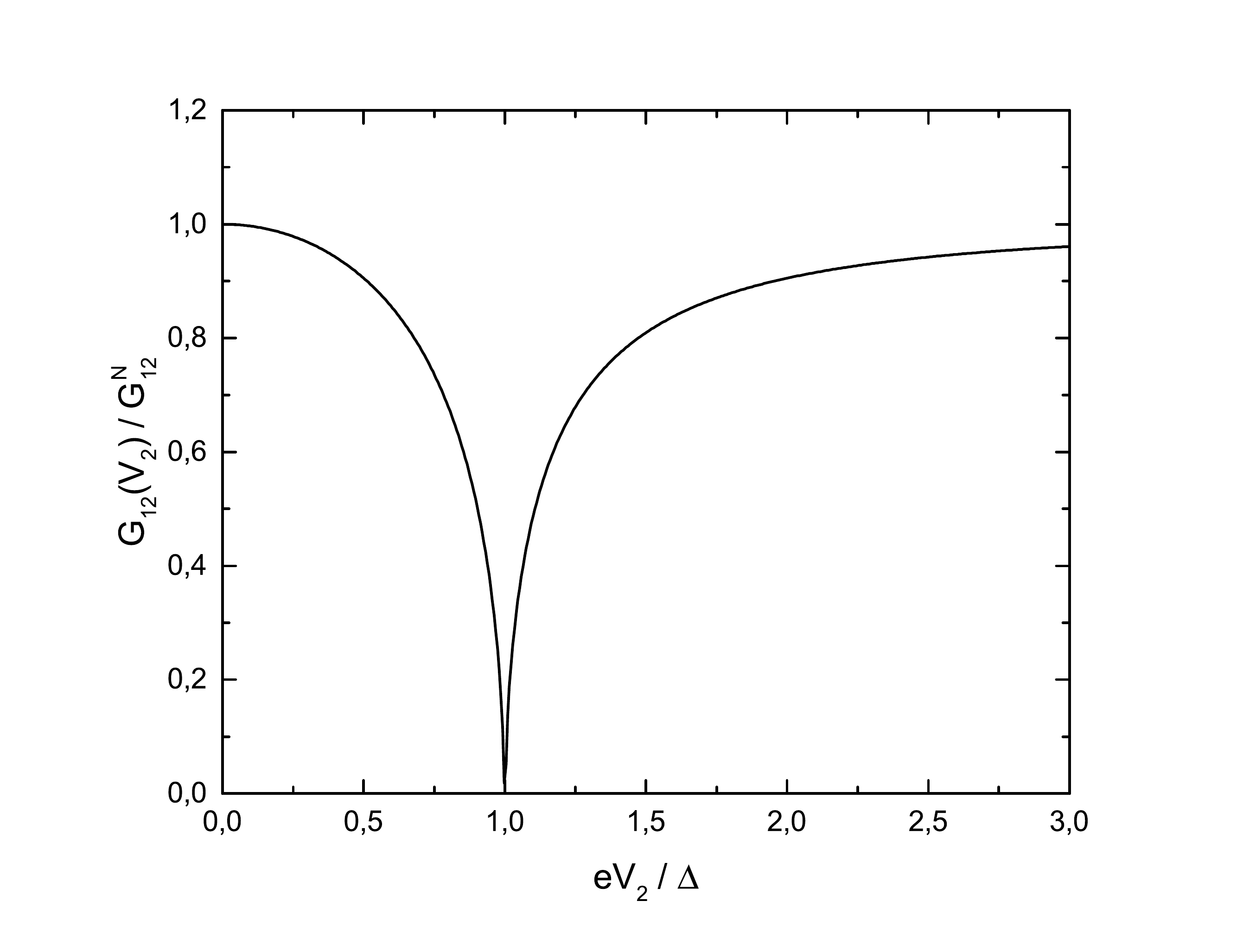} 
\caption{Zero temperature non-local conductance $\partial I_1/\partial V_2$ (determined by Eq. (\ref{G12diff}) with $E \to eV_2$)  as a function of the bias voltage $V_2$.}
\label{Fig:G12}
\end{figure}

In the tunneling limit $A_{1,n},A_{2,m}\ll 1$ one finds
\begin{eqnarray}
\gamma_+=\gamma_- \equiv \gamma_T = \frac{\sum_{n,m} \sqrt{A_{1,n}A_{2,m}}}{\sum_{n,m} A_{1,n}A_{2,m}}.
\end{eqnarray} 
Obviously in this regime we have $\gamma_\pm\gg 1$. Since in the other terms the prefactors are much smaller, 
one can keep only the terms $\propto \gamma_\pm$ in the expression (\ref{S12subgap}), thereby reproducing the result \cite{Hekking}
\begin{eqnarray}
S_{12} &=& \gamma_T G_{12}(0) \left[ e(V_1+V_2) \coth\frac{e(V_1+V_2)}{2T} 
\right.
\nonumber\\ &&
\left.
-\,  e(V_1-V_2) \coth\frac{e(V_1-V_2)}{2T} \right].
\label{S12tunnel}
\end{eqnarray}
The first and the second terms in the right-hand side of this formula are attributed respectively to CAR and EC processes. 
We observe that the noise cross-correlations remain positive, $S_{12}>0$, provided $V_1$ and $V_2$ have the same sign, 
and they turn negative, $S_{12}<0$, should $V_1$ and $V_2$ have different signs. 

In the opposite limit of perfectly conducting channels in both junctions with $\tau_{1,n}=\tau_{2,m}=1$ one gets
$\gamma_+=2$, $\gamma_-=0$, $\beta_1=\beta_2=0$. Hence, in this case we have \cite{GZ}
\begin{eqnarray}
S_{12} = G_{12}(0) \left[ -8T 
+ 2e(V_1+V_2) \coth\frac{e(V_1+V_2)}{2T}  \right].
\label{S12open}
\end{eqnarray}
This result is always positive at non-zero bias and low enough temperatures,
indicating the importance of CAR processes in this limit.
Note however, that in contrast to the tunnel limit (\ref{S12tunnel}), the last term in the Eq. (\ref{S12open}) is not necessarily proportional to the CAR probability.
Indeed, it may contain disorder averaged contributions of     
mixed processes involving both CAR and EC amplitudes \cite{Melin3,Ostrove}, originating from the general 
expression for the noise in terms of the  scattering matrix\cite{AD}.

Provided superconductivity gets totally suppressed (i.e. we set $\Delta\to 0$), it is straightforward to verify that our general expression for the cross-correlated noise  (\ref{S12full}) reduces to the result \cite{GZ2}
\begin{eqnarray}
S_{12} &=& -\frac{R_0^S}{R_1^NR_2^N}\bigg[ F_1 eV_1\coth\frac{eV_1}{2T} + F_2 eV_2\coth\frac{eV_2}{2T} 
\nonumber\\ &&
+\, (2-F_1-F_2) 2T\bigg],
\label{S12N}
\end{eqnarray}
where $F_j = \sum_n \tau_{j,n}(1-\tau_{j,n})/\sum_n \tau_{j,n}$ are the Fano factors of the contacts in the normal state.
We also note that in the large bias limit $eV_1,eV_2\gg \Delta$ Eq. (\ref{S12full})  equals to the normal state result (\ref{S12N}) plus voltage-independent  
excess noises related to both CAR and EC.

\begin{figure*}
\includegraphics[width=2\columnwidth]{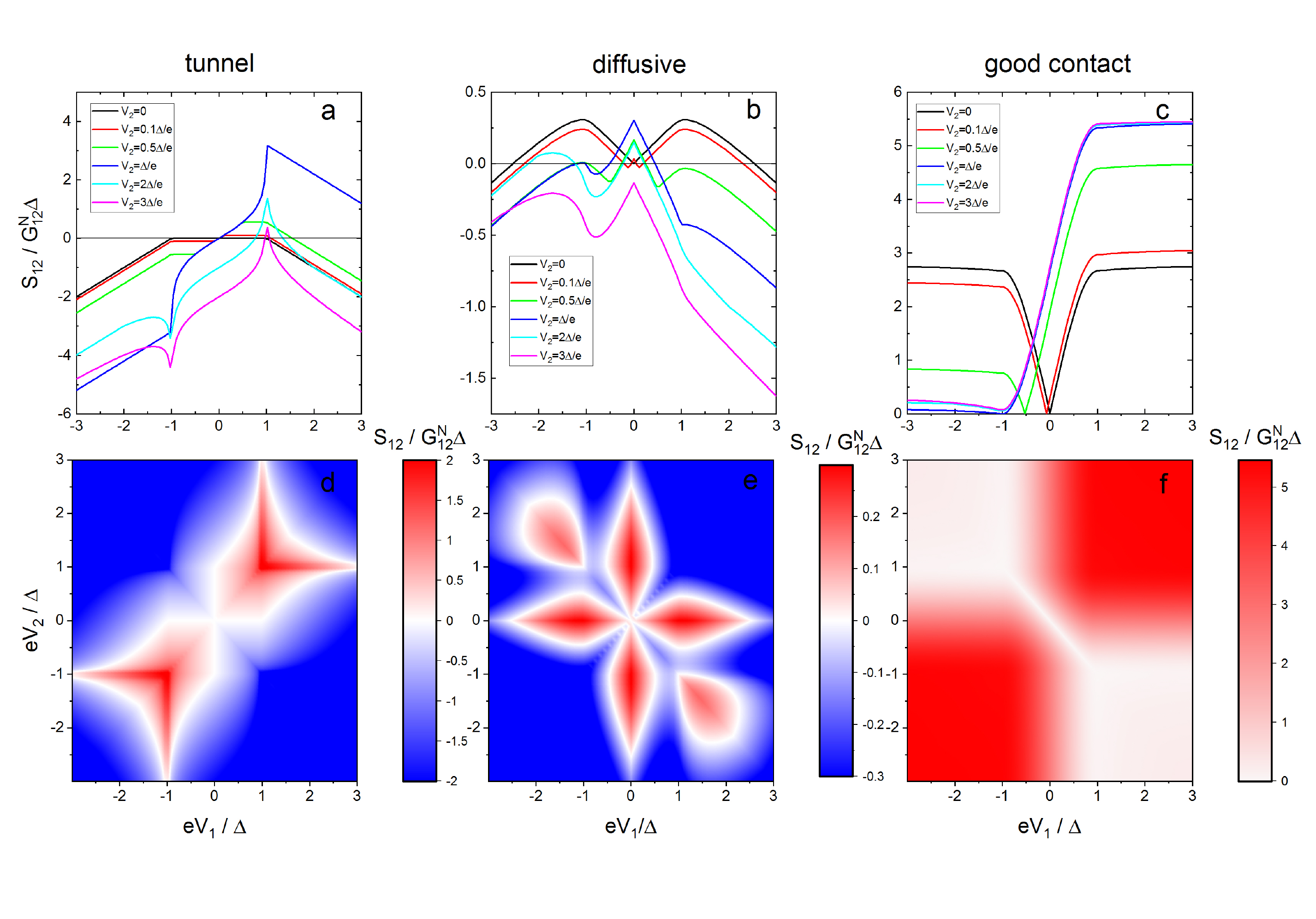}
\caption{Non-local shot noise in an NSN structure at zero temperature, $T=0$, with various types of contacts: 
(a,d) tunnel contacts, Eq. (\ref{S12full}) with $\tau_{1,n}=\tau_{2,m}= 0.1$; 
(b,e) diffusive contacts, Eq. (\ref{S12}); (c,f) fully open contacts, Eq. (\ref{S12full}) with $\tau_{1,n},\tau_{2,m}=1$ for all conducting channels. 
Graphs (a,b,c) show the dependence of the cross-correlated noise power $S_{12}$ on the bias voltage $V_1$ for several fixed values of $V_2$. Color plots (d,e,f) show the dependence of the noise cross correlations on both bias voltages $V_1,V_2$ with red and blue colors indicating respectively positive and negative cross correlations.  }
\label{Fig:3d}
\end{figure*}

Finally, let us analyze an important case of diffusive contacts.
Making use of Eq. (\ref{G1122}) and integrating over the transmission distribution  (\ref{Dorokhov}), we arrive at the non-local spectral conductance in the form
\begin{eqnarray}
&& G_{12}(E) = \frac{G_{12}^N}{4R_1^N R_2^N} 
\bigg[\theta(\Delta-|E|)\frac{\Delta^2-E^2}{E^2} 
\nonumber\\ &&
+\, \theta(|E|-\Delta)\frac{E^2-\Delta^2}{\Delta^2}  \bigg]
\left(\ln\left|\frac{|E|+\Delta}{|E|-\Delta}\right|\right)^2,
\label{G12diff}
\end{eqnarray}
where
$
G_{12}^N = R_0^S/(R_1^N R_2^N)
$
is the non-local conductance in the normal state.
This result can be easily derived by applying Kirchhoff's law to the equivalent circuit depicted in Fig. \ref{NISIN}b  
and assuming that $R_0^S\ll R_{1,2}^N$.
At $T=0$ the differential conductance $\partial I_1/\partial V_2$ exhibits the re-entrance effect (see Fig. \ref{Fig:G12}). 
Namely, one finds that $\partial I_1/\partial V_2|_{V_2=0}=\lim_{V_2\to\infty}\partial I_1/\partial V_2 = G_{12}^N$.

One can also work out a relatively simple analytical expression for  $S_{12}$. After averaging over the distributions (\ref{Dorokhov}), the expression 
(\ref{S12full}) reduces to the form
 \begin{widetext}
\begin{eqnarray}
S_{12} &=& G_{12}^N\int dE\big\{ -K_1(E/\Delta)\big[w(n_1^-,n_2^+)+w(n_1^+,n_2^-) - w(n_1^-,n_2^-) - w(n_1^+,n_2^+)\big]
\nonumber\\ &&
+\, K_2(E/\Delta) \big[w(n_1^-,n_2^+)+w(n_1^+,n_2^-) + w(n_1^-,n_2^-) + w(n_1^+,n_2^+)\big]  
 - K_3(E/\Delta)\big[ w(n_1^-,n_1^+) +  w(n_2^-,n_2^+)   \big]
\nonumber\\ &&
-\, K_4(E/\Delta) \big[w(n_1^-,n_S)+w(n_1^+,n_S) + w(n_2^-,n_S)+w(n_2^+,n_S) \big]
\nonumber\\ &&
-\, K_5(E/\Delta) \big[ w(n_1^-,n_1^-) + w(n_1^+,n_1^+)  + w(n_2^-,n_2^-) + w(n_2^+,n_2^+) \big] + 2K_4(E/\Delta) w(n_S,n_S) \big\}.
\label{S12}
\end{eqnarray}
\end{widetext}
Here we defined five dimensionless functions $K_j(x)$.
At $|x|<1$ (i.e. at $|E|<\Delta$) these functions read
\begin{eqnarray}
&& K_1 = x^2 \left(f_1-f_2\right)^2,\;\;
K_2 = (1-x^2)f_1^2/4,
\nonumber\\
&& K_3 = (1-x^2)\left(f_1-f_2\right)f_1,
\nonumber\\
&& K_4 = 0, \;\;
K_5 = (1-x^2)f_1f_2 / 2,
\nonumber
\end{eqnarray}
where the functions $f_1(x)$ and $f_2(x)$ are defined, respectively, in Eqs. (\ref{f1}) and (\ref{f2}).
For $|x|>1$  we have
\begin{eqnarray}
&& K_1 = \frac{\left(f_1-f_2\right)^2}{4x^2},\;
K_2 = -\frac{(x^2-1)f_2^2}{4x^4}, \; K_3 = 0,
\nonumber\\
&& K_4 = \frac{x^2-1}{2x^2}\left[f_1^2-2f_1f_2-\frac{f_2^2}{x^2}\right],\;
K_5 = \frac{(x^2-1)f_1f_2}{2x^2}.
\nonumber
\end{eqnarray}

The cross-correlated noise power for diffusive junctions (\ref{S12}) is plotted in Figs. \ref{Fig:3d}b and \ref{Fig:3d}e.
For comparison, in the same Figure we have also displayed the result (\ref{S12full}) in the tunneling limit (Figs. \ref{Fig:3d}a and \ref{Fig:3d}d) and 
for fully transparent junctions (Figs. \ref{Fig:3d}c and \ref{Fig:3d}f). For simplicity, in both these limiting cases we assume that all conducting
channels in the junctions have the same transparency ($\tau=0.1$ for the tunnel limit and $\tau=1$ for fully open NS junctions).
The dependence of $S_{12}$ on the bias voltage is asymmetric being very sensitive to the transparency of the junctions.
Curves of a similar shape have also been obtained numerically \cite{Melin2} for a ballistic NSN structure within the
scattering matrix approach\cite{AD}. Interestingly, for good contacts $S_{12}$ remains positive even for $e|V_{1,2}|>\Delta$, although at high bias
it becomes voltage independent, see Fig. \ref{Fig:3d}c.

In the limit of low voltages and temperatures $eV_{1,2},T\lesssim \Delta$ the energy integrals in Eq. (\ref{S12}) can be performed analytically, and we obtain
\begin{eqnarray}
S_{12} &=& G_{12}^N \bigg[ -\frac{16}{3}T - \frac{4}{3}eV_1\coth\frac{eV_1}{T} - \frac{4}{3}eV_2\coth\frac{eV_2}{T}
\nonumber\\ && 
+\, e(V_1+V_2)\coth\frac{e(V_1+V_2)}{2T}
\nonumber\\ && 
+\, e(V_1-V_2)\coth\frac{e(V_1-V_2)}{2T}\bigg].
\label{S12diffs}
\end{eqnarray}
Note that the last two terms in this expression for the cross-correlated current noise in NSN structures with diffusive contacts
resemble those of the result  (\ref{S12tunnel}) derived in the tunneling limit except the last term in Eq. (\ref{S12diffs}) enters with the opposite sign as compared to that in Eq. (\ref{S12tunnel}). The expression (\ref{S12diffs}) also follows from the general formula (\ref{S12subgap}), since for diffusive junctions one finds
$\beta_1=\beta_2=1/3$ and $\gamma_\pm = \pm 1$.
Depending on the bias voltages $V_1$ and $V_2$ the cross-correlated noise (\ref{S12diffs}) can take both positive and negative values, as it is illustrated in Fig. 4 b and e.
We also stress that the results for the non-local noise power derived here for the case of  diffusive contacts  cannot be correctly reconstructed  within a simple one-dimensional ballistic model \cite{Melin2}. Indeed, it is easy to check that, e.g., it order to get $\gamma_\pm=\pm 1$ within the latter model, for both barriers one should choose the same channel transmission value $\tau=2(\sqrt{2}-1)$. This choice, however, would then yield both subgap and overgap Fano factors, respectively $\beta_{1,2}$ and $F_{1,2}$, which do not correspond to the diffusive limit. Hence, e.g., the result in Eq. (\ref{S12diffs}) cannot be recovered from the model \cite{Melin2}.

In summary, we have developed a detailed theory describing both non-local electron transport and non-local shot noise
in three-terminal NSN hybrid structures with arbitrary distribution of transmissions for conducting channels in both NS junctions.
Our theory does not employ any restrictions imposed on the electron energy and, hence, remains applicable at all voltage bias values and at any temperature. In our analysis we paid particular attention to the physically important limit of diffusive NS junctions, in which case a non-trivial behavior of non-local shot noise  is recovered, exhibiting both
positive and negative cross correlations depending on the bias voltages. Our predictions allow to better understand the process of Cooper pair splitting in  NSN structures and are calling for their experimental verification.

\vspace{0.5cm}

\centerline{\bf Acknowledgements}

This work was supported in part by RFBR Grant No. 18-02-00586, and
by the Academy of Finland Centre of Excellence program (project 312057).

\appendix

\begin{widetext}

\section{}

Performing the averaging outlined in Sec. II we derive the local part of the CGF (\ref{Fr}) in the form
\begin{eqnarray}
&& {\cal F}_r(\chi_r) = {\rm tr}\,\ln\big[ \check I_{rr} -\check\Sigma_r(\chi_r) \langle\check G_{S,rr}\rangle \big]
= t\sum_n\int\frac{dE}{2\pi}\ln\big\{ 1+A_{r,n}W(2\chi_r,n_r^+,n_r^-)
\nonumber\\ &&
+\, (C_{r,n}+D_{r,n})\big[ W(\chi_r,n_r^+,n_S)+W(-\chi_r,n_r^-,n_S)\big]
+\, (C_{r,n}-D_{r,n})^2 W(\chi_r,n_r^+,n_S)W(-\chi_r,n_r^-,n_S)
\big\},
\label{Floc}
\end{eqnarray}
where
$
W(\chi,n_r,n_{r'}) = (e^{i\chi}-1)n_r(1-n_{r'}) + (e^{-i\chi}-1)(1-n_r)n_{r'}.
$
The CGF (\ref{Floc}) is equivalent to that derived in \cite{MK}.

The general expression for the cross-correlated current noise  $S_{12}$ which follows from our analysis in Sec. IV reads
\begin{eqnarray}
&& S_{12} = 
\frac{e^4R_0^S}{\pi^2} \sum_{n,m}\int_{|E|<\Delta} dE\, \frac{\Delta^2-E^2}{\Delta^2}A_{1,n}A_{2,m}\bigg\{
\bigg[\frac{(1-2A_{1,n})(1-2A_{2,m})}{\sqrt{A_{1,n}(0)A_{2,m}(0)}} 
- \frac{4E^2(1-A_{1,n})(1-A_{2,m})}{\Delta^2-E^2}\bigg]
\nonumber\\ && \times\,
\big[w(n_{1}^-,n_2^+)+w(n_{1}^+,n_2^-) - w(n_1^-,n_2^-)- w(n_1^+,n_2^+)\big]
\nonumber\\ &&
+\, w(n_1^-,n_2^+)+w(n_1^+,n_2^-)  + w(n_1^-,n_2^-)+w(n_1^+,n_2^+)  
- 4\big[ (1-A_{1,n}) w(n_1^-,n_1^+) + (1-A_{2,m}) w(n_2^-,n_2^+)   \big]
\nonumber\\ &&
-\, 2A_{1,n} \left( w(n_1^-,n_1^-) + w(n_1^+,n_1^+) \right) - 2A_{2,m}  \left(  w(n_2^-,n_2^-) + w(n_2^+,n_2^+) \right)
\bigg\}
\nonumber\\ &&
+\, \frac{e^4R_0^S}{\pi^2}\sum_{n,m}\int_{|E|>\Delta} dE\, \left(1-\frac{\Delta^2}{E^2}\right)g_{1,n}g_{2,m}
\bigg\{ - \frac{\Delta^2}{4(E^2-\Delta^2)}\left(1-\frac{g_{1,n}}{2}\right)\left(1-\frac{g_{2,m}}{2}\right)\big[ w(n_1^-,n_2^+) + w(n_1^+,n_2^-)
\nonumber\\ && 
-\, w(n_1^-,n_2^-) - w(n_1^+,n_2^+) \big]
- \frac{\Delta^2}{16E^2}g_{1,n}g_{2,m}\big[ w(n_1^-,n_2^+)+w(n_1^+,n_2^-)+w(n_1^-,n_2^-)+w(n_1^+,n_2^+) \big]
\nonumber\\ &&
+\,\left(\frac{\Delta^2}{8E^2}g_{1,n}g_{2,m} - \frac{1-g_{1,n}}{2}\right)\big[w(n_1^-,n_S)+w(n_1^+,n_S)-w(n_S,n_S)\big]
+\left( \frac{\Delta^2}{8E^2}g_{1,n}g_{2,m} - \frac{1-g_{2,m}}{2} \right)
\nonumber\\ && \times\,
\big[w(n_2^-,n_S)+w(n_2^+,n_S)-w(n_S,n_S)\big]
- \frac{g_{1,n}}{4} \big[ w(n_1^-,n_1^-) + w(n_1^+,n_1^+) \big]
-  \frac{g_{2,m}}{4} \big[ w(n_2^-,n_2^-) + w(n_2^+,n_2^+) \big]
\bigg\}.
\label{S12full}
\end{eqnarray}
\end{widetext}


\begin{thebibliography}{99}
\bibitem{Bel} W. Belzig, F.K. Wilhelm, C. Bruder, G. Sch\"on, and A.D. Zaikin, Superlatt. Microstruct. {\bf 25}, 1251 (1999).\bibitem{DF} G. Deutscher and D. Feinberg, Appl. Phys. Lett. {\bf 76}, 487 (2000).
\bibitem{Lesovik} G.B. Lesovik, T. Martin, and G. Blatter, Eur. Phys. J. B {\bf 24}, 287 (2001).
\bibitem{Buttiker} P. Samuelsson, E.V. Sukhorukov, and M. B\"uttiker, Phys. Rev. Lett. {\bf 91}, 157002 (2003).
\bibitem{Brange} F. Brange, O. Malkoc, and P. Samuelsson, Phys. Rev. Lett. {\bf 118}, 036804 (2017).
\bibitem{FFH} G. Falci, D. Feinberg, and F.W.J. Hekking, Europhys. Lett. {\bf 54}, 255 (2001).
\bibitem{BG} A. Brinkman and A.A. Golubov, Phys. Rev. B \textbf{74}, 214512 (2006).
\bibitem{KZ1} M.S. Kalenkov and A.D. Zaikin, Phys. Rev. B {\bf 75}, 172503 (2007).
\bibitem{KZ2} M.S. Kalenkov and A.D. Zaikin, Phys. Rev. B {\bf 76}, 224506 (2007).
\bibitem{Belzig} J.P. Morten, A. Brataas, and W. Belzig, Phys. Rev. B \textbf{74}, 214510 (2006).
\bibitem{GZ07} D.S. Golubev and A.D. Zaikin, Phys. Rev. B \textbf{76}, 184510 (2007).
\bibitem{LY} A. Levy Yeyati, F.S. Bergeret, A. Martin-Rodero, and T.M. Klapwijk, Nat. Phys. {\bf 3}, 455 (2007).
\bibitem{GKZ} D.S. Golubev, M.S. Kalenkov, and A.D. Zaikin, Phys. Rev. Lett. {\bf 103}, 067006 (2009). 
\bibitem{Beck1} D. Beckmann, H.B. Weber, and H. v. L\"ohneysen, Phys. Rev. Lett. {\bf 93}, 197003 (2004).
\bibitem{Teun} S. Russo, M. Kroug, T. M. Klapwijk, and A. F. Morpurgo, Phys. Rev. Lett. {\bf 95}, 027002 (2005).
\bibitem{Venkat1} P. Cadden-Zimansky and V. Chandrasekhar, Phys. Rev. Lett. {\bf 97}, 237003 (2006).
\bibitem{Hof} L. Hofstetter, S. Csonka, J. Nyg{\aa}rd, and C. Sch\"onenberger, Nature (London) {\bf 461}, 960 (2009).
\bibitem{Saclay} L. G. Herrmann, F. Portier, P. Roche, A. Levy Yeyati, T. Kontos, and C. Strunk, Phys. Rev. Lett. {\bf 104}, 026801 (2010).
\bibitem{Basel} A. Kleine, A. Baumgartner, J. Trbovic, D.S. Golubev, A.D. Zaikin, and C. Sch\"onenberger, 
Nanotechnology {\bf 21}, 274002 (2010).
\bibitem{Beck2} J. Brauer, F. H\"ubler, M. Smetanin, D. Beckmann, and H. v. L\"ohneysen, Phys. Rev. B {\bf 81}, 024515 (2010).
\bibitem{Basel3} J. Schindele, A. Baumgartner, and C. Sch\"onenberger, Phys. Rev. Lett. {\bf 109}, 157002 (2012).
\bibitem{Beck3} S. Kolenda, M.J. Wolf, D.S. Golubev, A.D. Zaikin, and D. Beckmann, Phys. Rev. B {\bf 88}, 174509 (2013).
\bibitem{BB} Ya.M. Blanter and M. B\"uttiker, Phys. Rep. {\bf 336}, 1 (2000).
\bibitem{Belzig3} J. B\"orlin, W. Belzig, and C. Bruder, Phys. Rev. Lett. {\bf 88}, 197001 (2002).
\bibitem{Hekking} G. Bignon, M. Houzet, F. Pistolesi, F. W. J.  Hekking,  Europhys. Lett. {\bf 67}, 110 (2004).
\bibitem{Samuelsson} P. Samuelsson and M. B\"uttiker, Phys. Rev. Lett. {\bf 89}, 046601 (2002).
\bibitem{Melin2} R. M\'elin, C. Benjamin, and T. Martin, Phys. Rev. B {\bf 77}, 094512 (2008).
\bibitem{Belzig2} J.P. Morten, D. Huertas-Hernando, W. Belzig, and A. Brataas, Phys. Rev. B {\bf 78}, 224515 (2008).
\bibitem{GZ} D.S. Golubev and A.D. Zaikin, Phys. Rev. B {\bf 82}, 134508 (2010).
\bibitem{Melin} A. Freyn, M. Fl\"oser, and R. M\'elin, Phys. Rev. B {\bf 82}, 014510 (2010).
\bibitem{Melin3} M. Fl\"oser, D. Feinberg, and R. M\'elin, Phys. Rev. B {\bf 88}, 094517 (2013).
\bibitem{Ostrove} C. Ostrove and L.E. Reichl, arXiv:1901.03766.
\bibitem{Ch} J. Wei and V. Chandrasekhar, Nat. Phys. {\bf 6}, 494 (2010).
\bibitem{Das} A. Das, Y. Ronen, M. Heiblum, D. Mahalu, A.V. Kretinin, and H. Shtrikman, Nat. Comm. {\bf 3}, 1165 (2012).
\bibitem{Carlos} J. C. Cuevas, A. Martin-Rodero, and A. Levy Yeyati, Phys. Rev. B {\bf 54}, 7366  (1996).
\bibitem{Volkov1} A.F. Volkov, Phys. Rev. Lett. {\bf 74}, 4730 (1995).
\bibitem{Morpurgo} J.J.A. Baselmans, A.F. Morpurgo, B.J. van Wees and T.M. Klapwijk, Nature {\bf 397}, 43 (1999).
\bibitem{Volkov2} R. Shaikhaidarov, A. F. Volkov, H. Takayanagi, V. T. Petrashov, and P. Delsing, Phys. Rev. B {\bf 62}, R14649 (2000).
\bibitem{Brouwer} I.L. Aleiner, P.W. Brouwer, and L.I. Glazman, Phys. Rep. {\bf 358}, 309 (2002).
\bibitem{BTK} G.E. Blonder, M. Tinkham, and T.M. Klapwijk, Phys. Rev. B {\bf 25}, 4515 (1982).
\bibitem{MK} B.A. Muzykantskii and D.E. Khmelnitskii, Phys. Rev. B {\bf 50}, 3982 (1994).
\bibitem{AD} M.P. Anantram and S. Datta, Phys. Rev. B {\bf 53}, 16390 (1996).
\bibitem{Dorokhov} O.N. Dorokhov, Solid State Comm. {\bf 51}, 381 (1984).
\bibitem{Beenakker} M.J.M. de Jong  and C.W.J. Beenakker,   Phys. Rev. B {\bf 49}, 16070 (1994).
\bibitem{Sanquer}  X. Jehl, M. Sanquer, R. Calemczuk, and D. Mailly, Nature {\bf 405}, 50 (2000).
\bibitem{Prober} A. A. Kozhevnikov, R. J. Schoelkopf, and D. E. Prober, Phys. Rev. Lett. {\bf 84}, 3398 (2000). 
\bibitem{HN1} F.W.J. Hekking and Yu.V. Nazarov, Phys. Rev. Lett. {\bf 71}, 1625 (1993).
\bibitem{HN2} F.W.J. Hekking and Yu.V. Nazarov, Phys. Rev. B {\bf 49}, 6847 (1994).
\bibitem{Tanaka} Y. Tanaka, A.A. Golubov, and S. Kashiwaya, Phys. Rev. B {\bf 68}, 054513 (2003).
\bibitem{GZ2} D.S. Golubev and A.D. Zaikin, Phys. Rev. B {\bf 85}, 125406 (2012).

\end{thebibliography}
\end{document}